\documentclass[nofootinbib,aps,twocolumn,round]{revtex4}
\usepackage{amsmath,amssymb}
\usepackage{dcolumn}
\usepackage{bm}
\usepackage{xcolor}
\usepackage{graphicx}
\usepackage{commath}
\usepackage{float}
\usepackage{hyperref}
\usepackage{esint}
\usepackage{mdsymbol}
\usepackage{booktabs}
\begin{document}

\title{Approximations for the divergence of the local large-scale structure velocity field and its implications for Tilted Cosmology}

\author{Erick Pastén$^2$}
\email{erick.contreras@postgrado.uv.cl}

\author{Sebastián Galvez$^3$}
\email{sebastian.galvez@usm.cl}

\author{V\'ictor H. C\'ardenas$^2$}
\email{victor.cardenas@uv.cl}

\affiliation{$^2$Instituto de F\'{\i}sica y Astronom\'ia, Universidad de
Valpara\'iso, Gran Breta\~na 1111, Valpara\'iso, Chile}

\affiliation{$^3$Centro científico tecnológico de Valparaíso, Universidad Federico Santa María, Av. España 1680, Valparaíso Chile}

\begin{abstract}

We characterize the peculiar velocity field of the local large-scale structure reconstructed from the $2M++$ survey, by treating it as a fluid, extracting the divergence via different approximations over a range pf averaged scales. This reconstructed field is important for cosmology, since it was used to correct the peculiar redshifts of the last SNIA compilation Pantheon+. The results have intriguing implications for the LLSS fluid dynamics and particularly for the ``Tilted Cosmology'' model, although those results have to be taken carefully as the velocity field could contain significant bias due to the reconstruction procedure. Those possible bias and its influence in our results are discussed. Representative values of the apparent deceleration parameter ($\Tilde{q}$) are computed, in order to compare our results with the theoretical predictions of the tilted-universe scenario. We conclude that better velocity field reconstructions are necessary in order to constrain the parameters implied in LLSS research and alternative cosmologies.

\end{abstract}


\maketitle


\section{Introduction}

Dark Energy (DE) is the usual explanation for the apparent universal acceleration implied by the SNIA data (\citep{Perlmutter99},\citep{Riess98}). However, the suggestion for the existence of dark energy is ultimately based on the cosmological principle, that is on the assumption of a globally homogeneous and isotropic Friedmann universe. The requirement of an extra parameter $\Omega_{\Lambda}$ is then necessary to explain the dimming of the supernovae magnitudes at large redshifts. Nevertheless, new interesting ideas have emerged in recent years, putting in doubt the cosmological principle, the Friedmann models and the existence of dark energy.

On sufficiently large scales the universe appears homogeneous and isotropic, according to the Cosmic Microwave Background (CMB) observations. On small scales, however, our cosmos is far from that, due to complex structures that produce overdensities/underdensities~\cite{Keenan13}, fractal-like structures \cite{Labini11,Labini98} and bulk peculiar motions that are not at rest with respect to the Hubble flow~\cite{Feindt13,Hudson99,Magoulas16}. There have been many works claiming that some of these effects can mimic an apparent acceleration. Possibly the combination of some (perhaps of all) of these contributions may have an effect stronger than we have previously thought~\cite{Celerier06,Enqvist07,Cosmai19,Tsagas2011,Asvesta22}. 

One of the proposed scenarios is the ``tilted cosmological model''~\cite{Tsagas2011}. The latter offers a natural environment for the theoretical study of the observed large-scale peculiar motions, by allowing for two groups of relatively moving observers. One group is aligned with the reference frame of the cosmos, which is identified with the coordinate system of the CMB, where the associated dipole vanishes by construction. The second group are the real observers, living in typical galaxies like our Milky Way and moving relative to the CMB frame (e.g.~see~\cite{Tsagas2008,Ellis2012}). Adopting a tilted almost-Friedmann universe and using linear relativistic cosmological perturbation theory, it was shown that relative-motion effects can lead to an apparent change in the sign of the deceleration parameter inside locally contracting bulk flows. Although the effect is a local artefact of the observers' peculiar motion, the affected scales can be large enough to have cosmological relevance. Then, observers inside (slightly) contracting bulk peculiar flows can be misled to believe that their universe recently entered a phase of accelerated expansion. Put another way, the unsuspecting observers may misinterpret the local contraction of the bulk flow they live in, as global acceleration of the surrounding universe (see~\cite{Tsagas2015,Tsagas2021,Tsagas2022} for further discussion and details). Our aim is to investigate this possibility by comparing theory to observations.  

We use the velocity field reconstruction from the 2M++ galaxy survey \cite{Carrick_2015}. This reconstruction provides a pair of data-cubes containing the density contrast and the velocity vectors in galactic coordinate. One can use these data to apply basic calculus and also to perform corrections due to peculiar velocities in cosmological data, as it was done in the last Pantheon+ SNIA compilation \cite{Scolnic_2022,Carr_2022}. In the present paper we estimate the average volume scalar of this local velocity-field reconstruction by different methods and on different scales. In all cases, the local bulk flow is found to contract on average, leading to negative values for the local deceleration parameter on a range of scales. These results seem to support the tilted cosmological scenario as an alternative natural explanation of the DE problem. However, the results have to be taken very carefully as the data could contain an important bias due to the reconstruction procedure.

In section~\ref{sTCM} we provide a brief but concise  description of tilted cosmological scenario, referring the reader to the related literature for more details. In section~\ref{cfa} we discuss how to relate the parameters obtained from the velocity-field reconstruction with the theory and in section~\ref{2MVF} we present the data used and discuss the impact of the possible bias in the data. Finally, we summarize the method and the results obtained from this reconstruction in sections~\ref{Meth} and \ref{Res} and discuss their implications for cosmology at the end of this paper. In addition to cosmology, our analysis has potential applications to astrophysics and to the local  structure dynamics.

\section{Tilted cosmology model}\label{sTCM}
Consider a perturbed Friedmann-Robertson-Walker (FRW) universe with two groups of relatively moving observers. Assuming that $u_a$ and $\tilde{u}_a$ are the 4-velocities of these observers and $v_a$ is the (non-relativistic) peculiar velocity of the latter group with respect to the former, we have
\begin{equation}
\tilde{u}_a= u_a+ v_a\,,  \label{4vels}
\end{equation}
to first approximation (with $u_av^a=0$ always). Introducing two sets of observers means that (strictly speaking) there are two temporal directions (along $u_a$ and $\tilde{u}_a$) and two associated 3-spaces (orthogonal $u_a$ to and $\tilde{u}_a$). Then, the corresponding (covariant) differential operators are $\dot{}= u^a\nabla_a$ and ${}^{\prime}=\tilde{u}^a\nabla_a$ for the time derivatives, with ${\rm D}_a=h_a{}^b\nabla_b$ and $\tilde{\rm D}_a=\tilde{h}_a{}^b\nabla_b$ for the spatial gradients. Also, the tensors $h_{ab}=g_{ab}+u_au_b$ and $\tilde{h}_{ab}=g_{ab}+ \tilde{u}_a\tilde{u}_b$ project orthogonal to $u_a$ and $\tilde{u}_a$ respectively. 

The kinematic information of the observers' motion is decoded by decomposing the gradient of their 4-velocity field as follows
\begin{equation}
\nabla_bu_a= \frac{1}{3}\,\Theta h_{ab}+ \sigma_{ab}+ \omega_{ab}- A_au_b\,.  \label{Nbua}
\end{equation}
In the above, $\Theta$ is the volume expansion/contraction scalar (when positive/negative respectively), $\sigma_{ab}$ is the shear, $\omega_{ab}$ is the vorticity and $A_a$ is the 4-acceleration (e.g.~see~\cite{Tsagas2008,Ellis2012}). In an exactly analogous way, the $\tilde{u}_a$-field splits as $\nabla_b\tilde{u}_a=(\tilde{\Theta}/3)\tilde{h}_{ab}+ \tilde{\sigma}_{ab}+\tilde{\omega}_{ab}- \tilde{A}_a\tilde{u}_b$, with the tildas denoting variables evaluated in the tilted frame of the bulk flow. Relative to the same coordinate system, the peculiar-velocity field splits as
\begin{equation}
\tilde{\rm D}_b\tilde{v}_a=  \frac{1}{3}\,\tilde{\theta}\tilde{h}_{ab}+ \tilde{\varsigma}_{ab}+ \tilde{\varpi}_{ab}\,,  \label{tDntva}
\end{equation}
where $\tilde{\theta}$, $\tilde{\varsigma}_{ab}$ and $\tilde{\varpi}_{ab}$ are the volume scalar, the shear and the vorticity of the bulk peculiar motion~\cite{Tsagas2015}. Of the last three variables, the most important for our purposes is the peculiar volume scalar ($\tilde{\theta}$), which takes positive/negative values in locally expanding/contracting bulk flows respectively.

The three kinematic sets defined above are related by lengthy nonlinear expressions (e.g.~see~\cite{Maartens1998} for the full list). Assuming non-relativistic peculiar motions on an FRW background, we obtain the linear relations
\begin{equation}
\Tilde{\Theta}= \Theta+ \tilde{\theta} \hspace{10mm} {\rm and} \hspace{10mm} \Tilde{\Theta}^{\prime}= \dot{\Theta}+ \tilde{\theta}^{\prime}\,,  \label{Thetas}
\end{equation}
between the volume scalars and between their time derivatives evaluated in the two frames. At this point, we note that $\Theta$ and $\tilde{\Theta}$ monitor the expansion rate of the universe, namely the Hubble parameters, as measured in their corresponding frames (that is $\Theta=3H$ and $\tilde{\Theta}=3\tilde{H}$). Then, equations (\ref{Thetas}a) and (\ref{Thetas}b) imply that the expansion and the acceleration/deceleration rates measured in the tilted coordinate system differ from those measured in its CMB counterpart solely due to relative-motion effects. In particular, recalling that
\begin{equation}
q= -1- {3\dot{\Theta}\over\Theta^2} \hspace{10mm} {\rm and} \hspace{10mm} \tilde{q}= -1-{3\tilde{\Theta}^{\prime}\over\tilde{\Theta}^2}\,,  \label{qs} 
\end{equation}
define the deceleration parameters in the CMB and the bulk-flow frames respectively, the following useful relation between $\tilde{q}$ and $q$ can be obtained \cite{Tsagas2015,Tsagas2021}:
\begin{equation}
\tilde{q}= q+ {\tilde{\theta}^{\prime}\over2\dot{H}}\,,  \label{tq1}
\end{equation}
to first approximation. Recall that $\tilde{\Theta}=\Theta=3H$ in the Friedmann background. Also note that, whereas $\tilde{\theta}/H\ll1$ at the linear level, the ratio $\tilde{\theta}^{\prime}/\dot{H}$ of their time derivatives is not always small. Finally, using relativistic linear cosmological perturbation theory, we arrive at:
\begin{equation}
\tilde{q}= q+ {1\over9}\left({\lambda_H\over\lambda}\right)^2 {\tilde{\theta}\over H}\,.  \label{tq2}
\end{equation}
with $\lambda_H=1/H$ and $\lambda$ representing the Hubble horizon and the scale of the bulk flow in question. Note that we have focused on bulk peculiar flows with sizes considerably smaller than the Hubble length (i.e.~$\lambda\ll\lambda_H$ -- see~\cite{Tsagas2015,Tsagas2021} for the full details of the derivation).\footnote{Expression (\ref{tq2}) has been obtained on an Einstein-de Sitter background, primarily for reasons of mathematical simplicity. It is fairly straightforward to show that the linear result (\ref{tq2}) holds on essentially all FRW backgrounds, irrespective of their equation of state and spatial curvature~\cite{Tsagas2022}.}

Following (\ref{tq2}), the deceleration parameter measured locally by the bulk flow observers ($\tilde{q}$) differs from that of the global universe, which by definition coincides with the deceleration parameter measured in the idealised CMB frame ($q$). The difference is entirely due to the peculiar motion of the tilted observer, since $\tilde{q}=q$ when $\tilde{\theta}=0$. Also, the ``correction'' term in (\ref{tq2}) is scale-dependent and it gets stronger on progressively smaller scales (i.e.~for $\lambda\ll\lambda_H$), despite the fact that $\tilde{\theta}/H\ll1$ throughout the linear regime. Moreover, in accord with (\ref{tq2}), the overall impact of relative motion on $\tilde{q}$ is also determined by the sign of the peculiar volume scalar ($\tilde{\theta}$). The latter is positive in locally expanding bulk flows, which means that the deceleration parameter measured by observers residing in them will be larger than that of the actual universe (i.e.~$\tilde{q}>q$ when $\tilde{\theta}>0$). In the opposite case, that is inside locally contracting bulk flows, the local deceleration parameter becomes smaller (i.e.~$\tilde{q}<q$ for $\theta<0$). The latter case is clearly the most intriguing, since it allows for the sign of the deceleration parameter to change, from positive to negative, when measured by observers inside locally contracting bulk flows. Although the sign-change of $\tilde{q}$ is simply an illusion and a local artefact of the observer's relative motion, the affected scales can be large enough to make it look as a recent global event. If so, an unsuspecting observer may be misled to believe that their universe has recently entered a phase of accelerated expansion. According to (\ref{tq2}), the ``transition scale'', where the local deceleration parameter crosses the $\tilde{q}=0$ threshold is~\cite{Tsagas2021} 
\begin{equation}
\lambda_T= {1\over3}\,\sqrt{\tilde{|\theta}|\over qH}\,\lambda_H\,,  \label{lambdaT}
\end{equation}
where $q>0$ always (with $q=1/2$ in the case of the Einstein-de Sitter background).

Although theoretically the model outlined above is well developed, it is not obvious yet how one should relate the tilted cosmological scenario to the observations. Parametrizing the deceleration function as $\tilde{q}=\tilde{q}(z)$ and then using it in (\ref{tq2}), has led to a good fit with the Pantheon SNIA sample~\cite{Asvesta22}. Also, an apparent (Doppler-like) dipole anisotropy is expected to appear in the observed distribution of the local deceleration parameter ($\tilde{q}$), due to the bulk-flow motion relative to the CMB frame~\cite{Tsagas2011}. However, which observational frame (heliocentric, geocentric, galactic, or cosmological) should be employed and what peculiar-velocity corrections should be applied to the data, in order to observe the aforementioned dipolar anisotropy, are the subjects of ongoing debate~\cite{Colin19,Rubin_2020,ColinResponse}. In this paper, our aim is to study the dynamical structure of the peculiar velocity field directly from data reconstruction. We choose the velocity field reconstruction of the 2M++ survey \cite{Carrick_2015}, which has been previously used in cosmology to correct the peculiar velocities of the SNIA data in the last Pantheon+ compilation~\cite{Scolnic_2022,Carr_2022}. The methods used to characterize this peculiar velocity field and the procedures employed to relate our results with those of the tilted cosmologies are discussed in the next sections.

\section{Classical Fluid approximation}\label{cfa}
The kinematic analysis outlined in the previous section, is straightforwardly adapted to the Newtonian framework as well (e.g.~see~\cite{Ellis1971,Ellis1990} for further discussion and details). In so doing, one replaces the projector ($h_{ab}$), which also acts as the metric tensor of the 3-space, with the Kronecker delta ($\delta_{ij}$). Also, time derivatives and 3-dimensional covariant gradients are replaced by convective derivatives and by ordinary partial derivatives respectively. Note that, given the near spatial flatness of the observed universe, any curvature corrections due to a nonzero connection ($\Gamma^a{}_{bc}$) will be of the second perturbative order. Then, focusing on the peculiar-velocity field ($\mathbf{v}=v_i$), we have 
${\tilde{\theta}}_{ij}=\nabla\mathbf{v}=\partial_jv_i$ and
\begin{eqnarray}
    \tilde{\theta}_{ij}= \frac{1}{3}\, \tilde{\theta}\delta_{ij}+\tilde{\varsigma}_{ij}+ \tilde{\varpi}_{ij}\,.
\end{eqnarray}
Here, the (local) volume expansion/contraction scalar, the shear tensor and the vorticity tensor of the bulk peculiar flow are respectively defined as
\begin{eqnarray}
    \tilde{\theta}&=&\partial_iv_i= \delta^{ij}\partial_jv_i\,, \\
    \tilde{\varsigma}_{ij}&=&\frac{1}{2} \left(\partial_j v_i+\partial_iv_j\right)- \frac{1}{3}\,\tilde{\theta}\,\delta_{ij}, \\
    \tilde{\varpi}_{ij}&=&\frac{1}{2} \left(\partial_j v_i- \partial_iv_j\right)\,.
\end{eqnarray}

It ts possible to evaluate the gradient tensor of the local peculiar-velocity field using this approach. In particular, the gradient tensor can be reduced to the $3\times3$-matrix of the partial derivatives of the $\tilde{v}_i$-field as:

\begin{eqnarray}
    \tilde{\theta}_{ij}=\partial_j v_i=\begin{pmatrix}
        \frac{\partial v_x}{\partial x} & \frac{\partial v_x}{\partial y} & \frac{\partial v_x}{\partial y} \\
        \frac{\partial v_y}{\partial x} & \frac{\partial v_y}{\partial y} & \frac{\partial v_y}{\partial y} \\
        \frac{\partial v_z}{\partial x} & \frac{\partial v_z}{\partial y} & \frac{\partial v_z}{\partial y} \\
    \end{pmatrix}\,,
\end{eqnarray}
directly relating $\tilde{\theta}_{ij}$ to the Jacobian tensor of the field. 



\begin{figure}[ht]
    \centering
    \includegraphics[width=8cm]{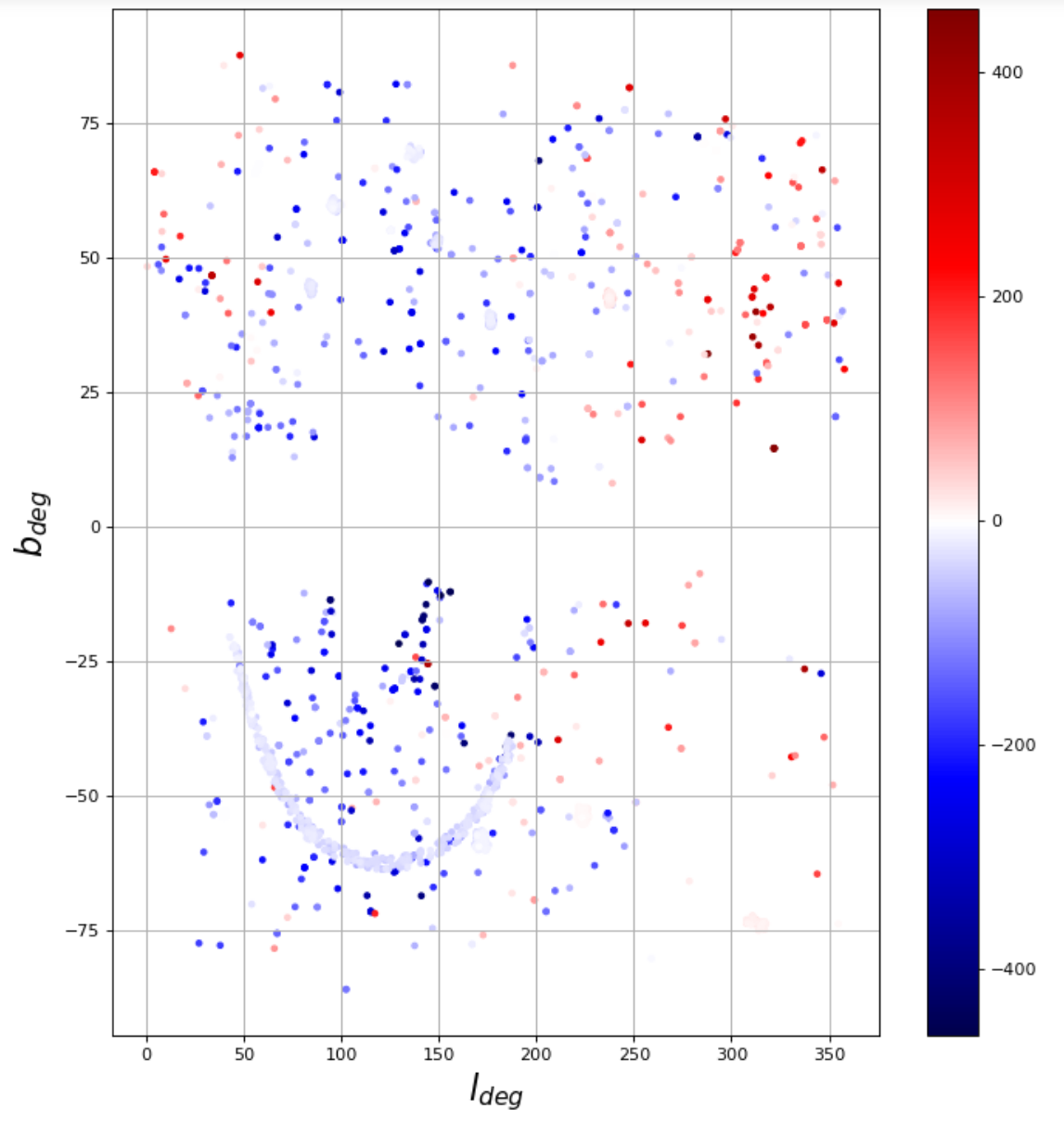}
    \caption{Peculiar velocities of the Pantheon+ SNIA compilation}
    \label{PVSNIA}
\end{figure}

\begin{figure*}[ht]
    \centering
    \includegraphics[width=8cm]{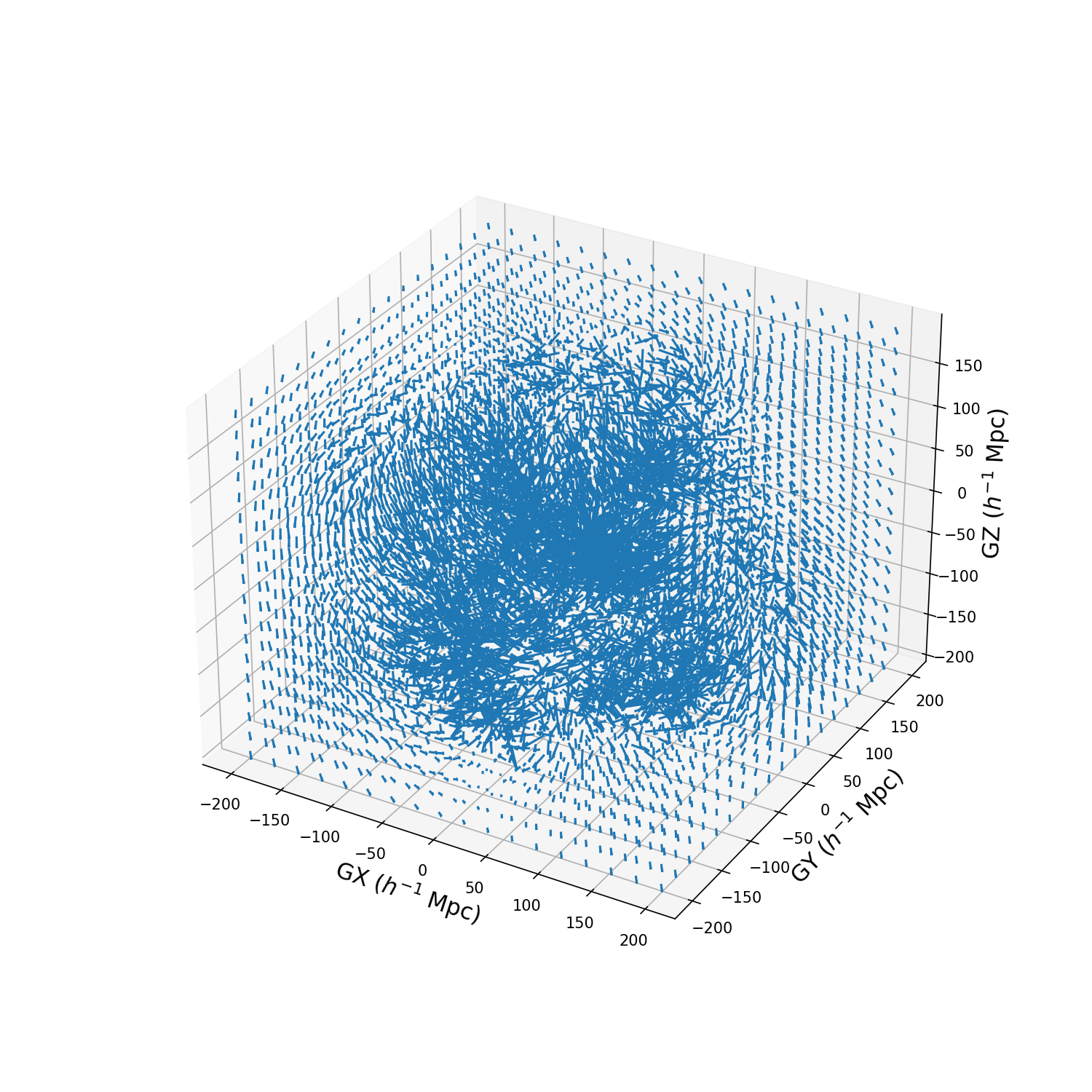}\includegraphics[width=8cm]{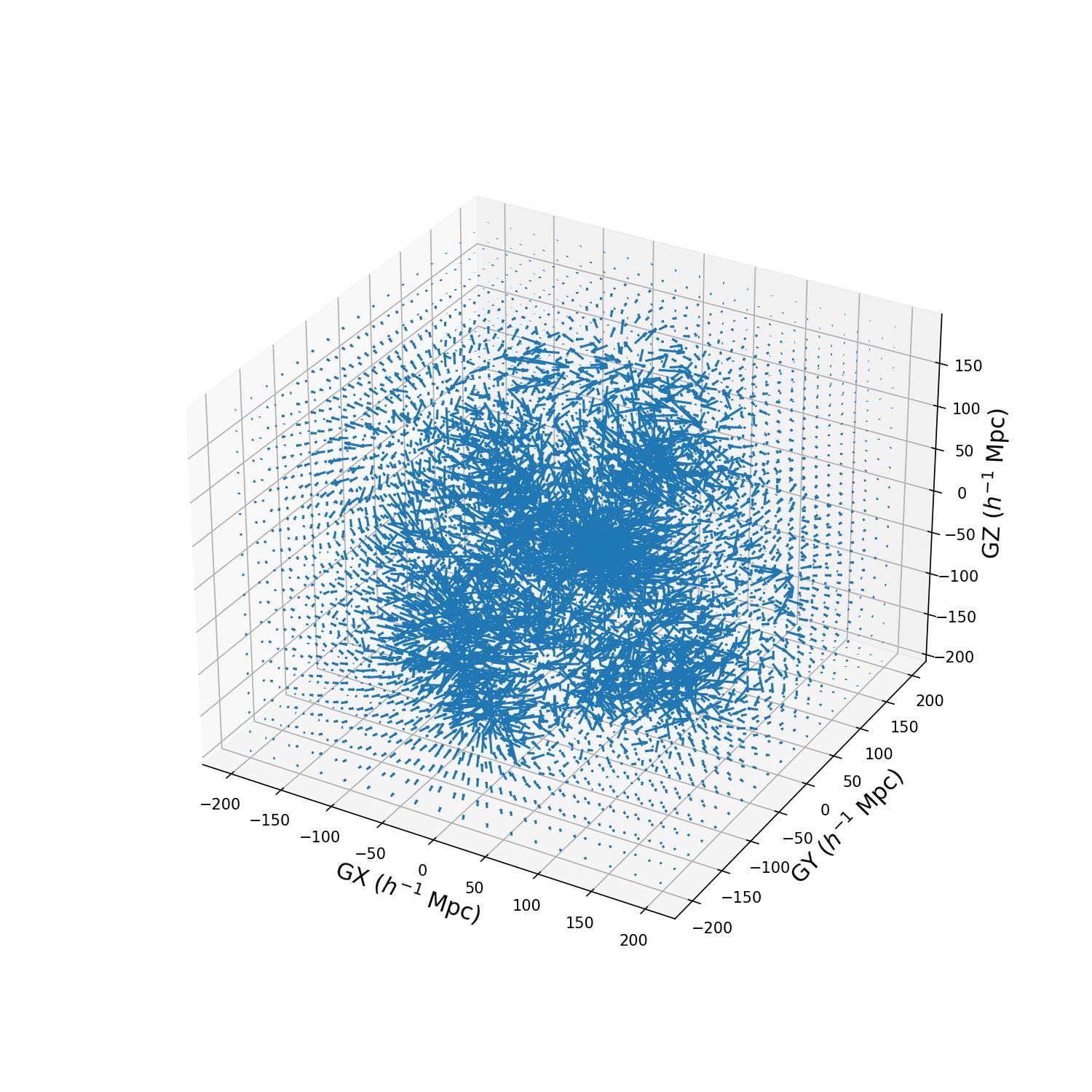}
    \caption{Peculiar Velocity field reconstruction from the 2M++ density field in galactic coordinates. Visualizations in 3D, with (left) and without (right) external dipole component.}
    \label{PV3D}
\end{figure*}

\section{The 2M++ velocity field reconstruction}\label{2MVF}

The last Supernovae IA compilation, namely Pantheon+~\cite{Scolnic_2022}, was released in 2022 showing a great improvement in the utility of data at low redshifts for cosmological uses. Part of this improvement is due to better corrections of the peculiar velocities of the SNIA data~\cite{Carr_2022} (see Figure~\ref{PVSNIA}). This was done by using a velocity field reconstruction based on the 2M++ galaxy survey \cite{Carrick_2015} (see Figure~\ref{PV3D}). The reconstruction procedure can be summarized as follows. If $\delta(\mathbf{r})$ is the density contrast, then the peculiar velocity field can be approximated as proportional to the gravitational acceleration when the fluctuations are small:

\begin{equation}
    \mathbf{v}(\mathbf{r})=\frac{f(\Omega_m)}{4\pi}\int d^3\mathbf{r'}\delta(\mathbf{r'})\frac{\mathbf{r'}-\mathbf{r}}{     \lvert \mathbf{r'}-\mathbf{r}\rvert^3 }\,. \label{Reconstruction}
\end{equation}
Here, $f$ is the growth rate of cosmic structures defined as $f=\Omega_m^\gamma$, where $\gamma=0.5$ for $\Lambda$CDM cosmology. Also, $r=HR$ is measured in $km/s$ where $R$, with $R$ being the comoving distance in $Mpc$ and $H$ the Hubble parameter. 

Since the total density perturbation ($\delta$) cannot be directly observed, a bias parameter ($b$) has been introduced to relate the observed density contrast ($\delta_g$) with the real one:
\begin{equation}
    \delta=\frac{\delta_g}{b}\,,
\end{equation}
at the linear level. Therefore, the important parameter in evaluating the velocity field is the ratio $\beta=f/b$, since we can write:
\begin{equation}
    \mathbf{v}(\mathbf{r})=\frac{\beta}{4\pi}\int d^3\mathbf{r'}\delta_g(\mathbf{r'})\frac{\mathbf{r'}-\mathbf{r}}{     \lvert \mathbf{r'}-\mathbf{r}\rvert^3 }\,,
\end{equation}
to relate directly the peculiar velocity with the observed galaxy density. Also, as the observations extend only up to a maximum scale ($R_{max}$), the contribution beyond this length can be added as a constant external velocity parameter ($\mathbf{V}_{ext}$), so that finally:
\begin{equation}
    \mathbf{v}(\mathbf{r})=\frac{\beta}{4\pi}\int^{R_{max}}d^3\mathbf{r'}\delta_g(\mathbf{r'})\frac{\mathbf{r'}-\mathbf{r}}{     \lvert \mathbf{r'}-\mathbf{r}\rvert^3 }+\mathbf{V}_{ext}\,,  \label{velrec}
\end{equation}
where $\beta$ and $\mathbf{V}_{ext}$ are determined empirically from the reconstruction of the density field.

\begin{figure*}[ht]
    \centering
    \includegraphics[width=8cm]{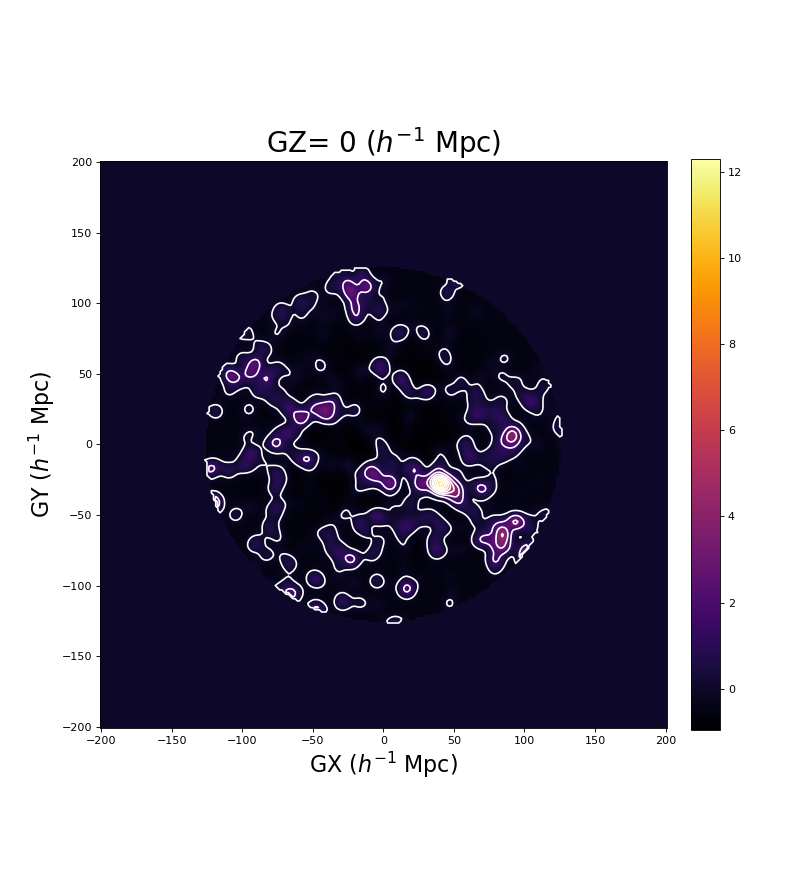}
    \includegraphics[width=8cm]{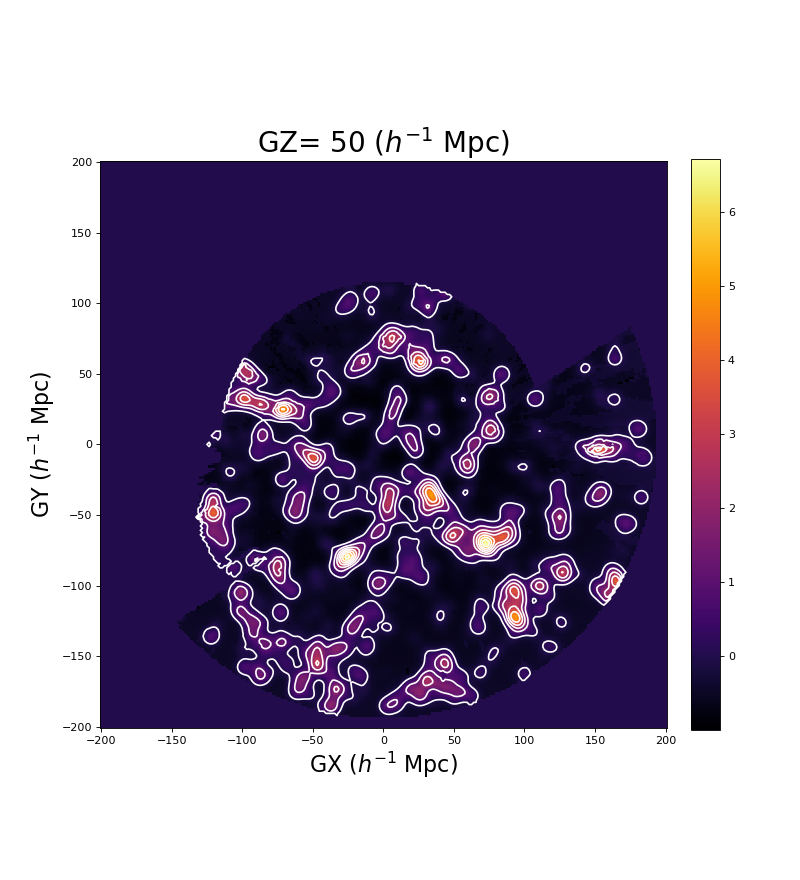}
    \includegraphics[width=8cm]{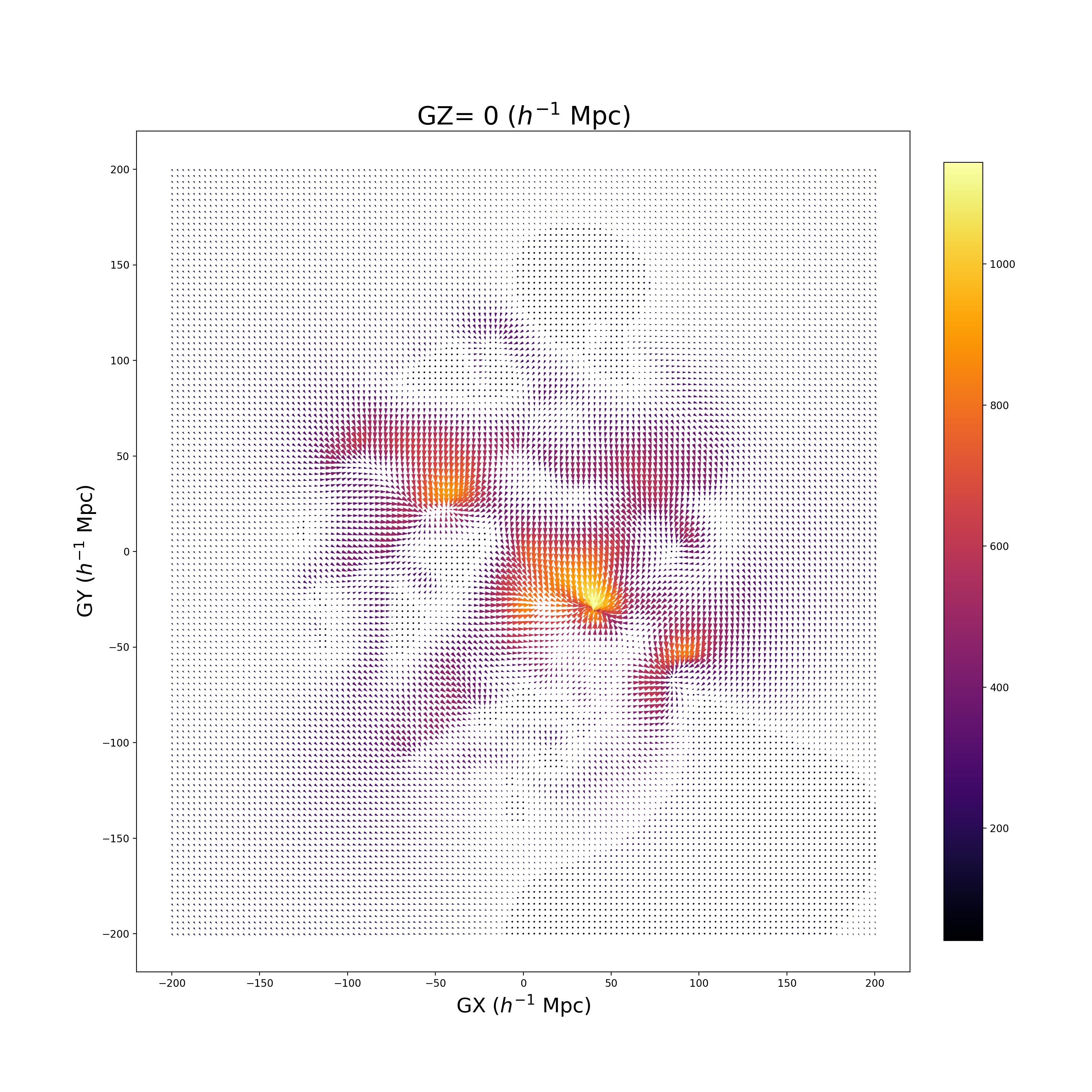}\includegraphics[width=8cm]{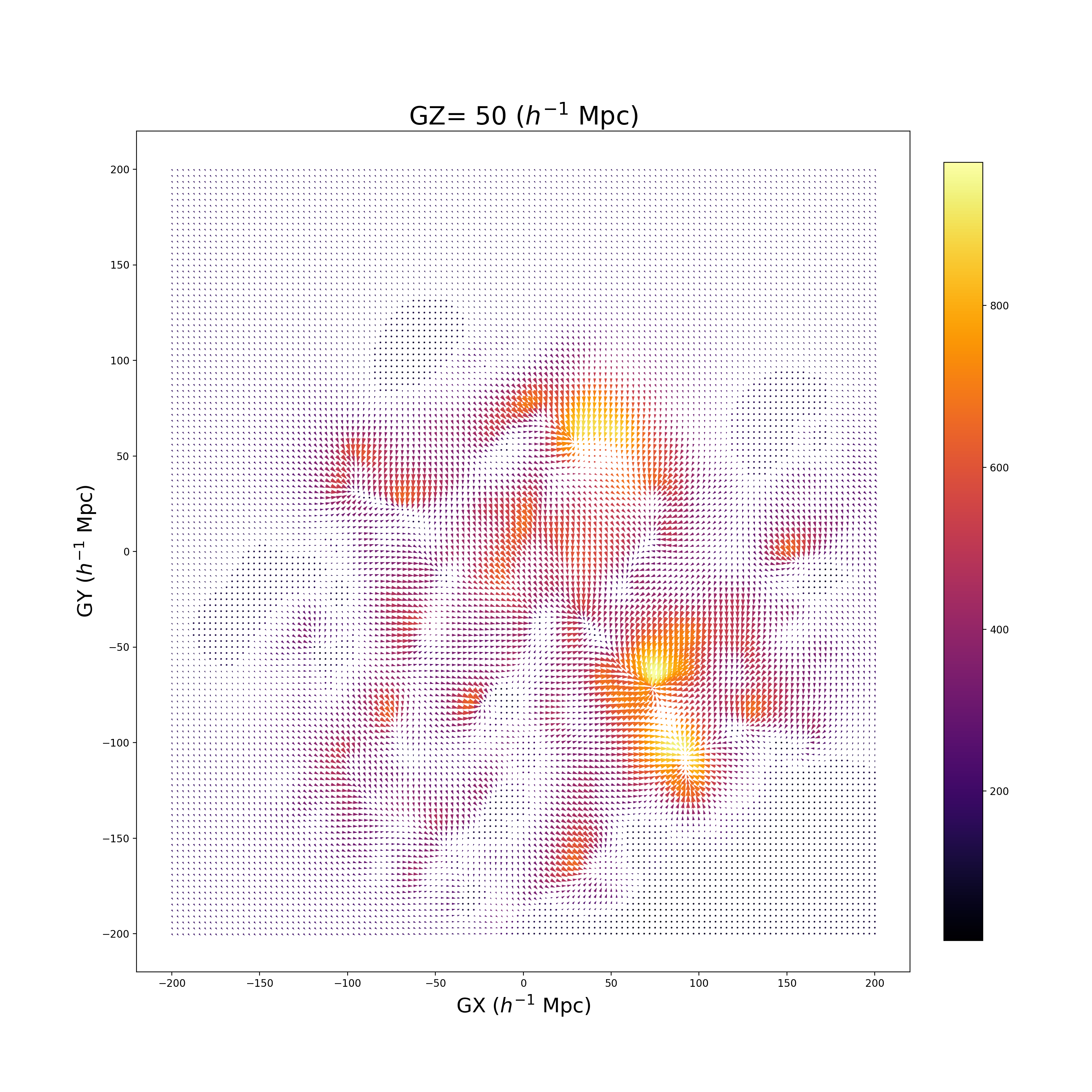}
    \caption{The density contrast and the vector velocity field projected over the $GZ=0$ and $GZ=50 h^{-1}Mpc$ galactic planes.}
    \label{DensityVel}
\end{figure*}

We use the density contrast and the velocity field (see Figure~\ref{DensityVel})) given by~\cite{Carrick_2015}, which can be easily downloaded from \href{https://cosmicflows.iap.fr/}{https://cosmicflows.iap.fr/}. There, the authors provide two useful data-cubes containing the density contrast $\delta$ and the velocity field $\mathbf{v}$ using the best-fit parameters $\beta=0.431 \pm 0.021$ and $\mathbf{V}_{ext}=(89\pm21,-131\pm23,17\pm26)\>km/s$ (with $\rvert \mathbf{V}_{ext}\lvert=159\pm23\>km/s$) in galactic Cartesian coordinates. It is also important to note that a different value of $\beta$, namely $\beta=0.341 \substack{+0.031 \\ -0.047}$, was used to correct the Pantheon+ data~\cite{Said_20,Carr_2022}, claiming that it gives a better fit when comparing the SDSS Fundamental Plane peculiar velocities to the predicted peculiar velocity field. Overall, we can write $\mathbf{v}=\beta\mathbf{v}_{rec}$, where $\mathbf{v}_{rec}$ gives the directions and relative magnitudes of the velocity field. Then, it is easy to use both values and compare the results.

In order to apply the same corrections to Pantheon+, the whole velocity field was approximated by a radially decaying function along the direction of the bulk flow. The latter is a 200~$Mpc$ sphere, composed by the sum of an external $\mathbf{V}_{ext}$ and a small average internal velocity $\mathbf{v}_{200}$. Interestingly, the external dipole component does not contribute to the gradient as it is a constant.\footnote{Even a radially decaying $\mathbf{V}_{ext}$ function, with a fixed direction, does not affect the average divergence of the velocity field.} Therefore:

\begin{equation}
    \nabla  \mathbf{v}= \nabla  (\beta \mathbf{v}_{rec} + \mathbf{V}_{ext})= \beta \nabla \mathbf{v}_{rec}\,.
\end{equation} 

\subsection{Bias in the velocity field reconstruction} 

The divergence of the reconstructed velocity field could be computed directly as:

\begin{equation*}
    \nabla \cdot \mathbf{v}(\mathbf{r})=-\beta \delta_g(\mathbf{r})\,.
\end{equation*}

This is expected as in linear theory; the divergence of the velocity field is proportional to the density contrast. However, is important to note that according to \cite{Carrick_2015} this density contrast is normalized over the mean density of a selected scale of the survey denoted by $\bar{\rho}$. Therefore, this could introduce an important bias in the velocity field. To analyse this, we define $\delta$ as the real density contrast:

\begin{equation*}
    \delta=\frac{\rho-\rho_0}{\rho_0},
\end{equation*}

Here, $\rho_0$ refers to the mean density of the universe. It is possible that this density contrast, denoted by $\delta$, is different from the density contrast $\delta_s$ of the reconstruction:

\begin{equation*}
    \delta_s=\frac{\rho-\bar{\rho}}{\bar{\rho}},
\end{equation*}

where $\bar{\rho}$ refers to the mean density of the survey area considered (in this case, over a scale of $200 \frac{Mpc}{h}$ or $z \sim 0.07$). Note that, according to this definition, the averaged density contrast over the entire scale of the survey, and therefore the averaged divergence of the velocity field, should be zero. If we eliminate the real density $\rho$ from the two equations, we can obtain the relation:

\begin{eqnarray}\label{bias}
    (\bar{\delta}+1)\delta_s+\bar{\delta}=\delta
\end{eqnarray}

Where $\bar{\delta}$ is the mean density contrast of the $200 \frac{Mpc}{h}$ scale w.r.t. to the density of the universe:

\begin{eqnarray}
    \bar{\delta}=\frac{\bar{\rho}-\rho_0}{\rho_0}
\end{eqnarray}

Therefore, $\bar{\delta}$ quantify the deviation that $\delta_s$ has from the real value $\delta$. We will consider this possible effect in our analysis.

Overall, is important to note that according to \cite{Carrick_2015} in section 5.3.4, the authors state that the average density contrast over different regions of the survey is consistent with some observational results in \cite{Whitbourn2014,Bohringer2014}. Additionally, it is worth remembering that this velocity field has already been used to correct peculiar velocities in the last Pantheon+ compilation, which is crucial for observational cosmology. Therefore, we conclude that while there could be a bias in the data that we need to consider carefully, this reconstruction field is a good starting point for the extraction of interesting parameters for LLSS research.

\begin{figure*}[ht]
    \centering
    \includegraphics[width=15cm]{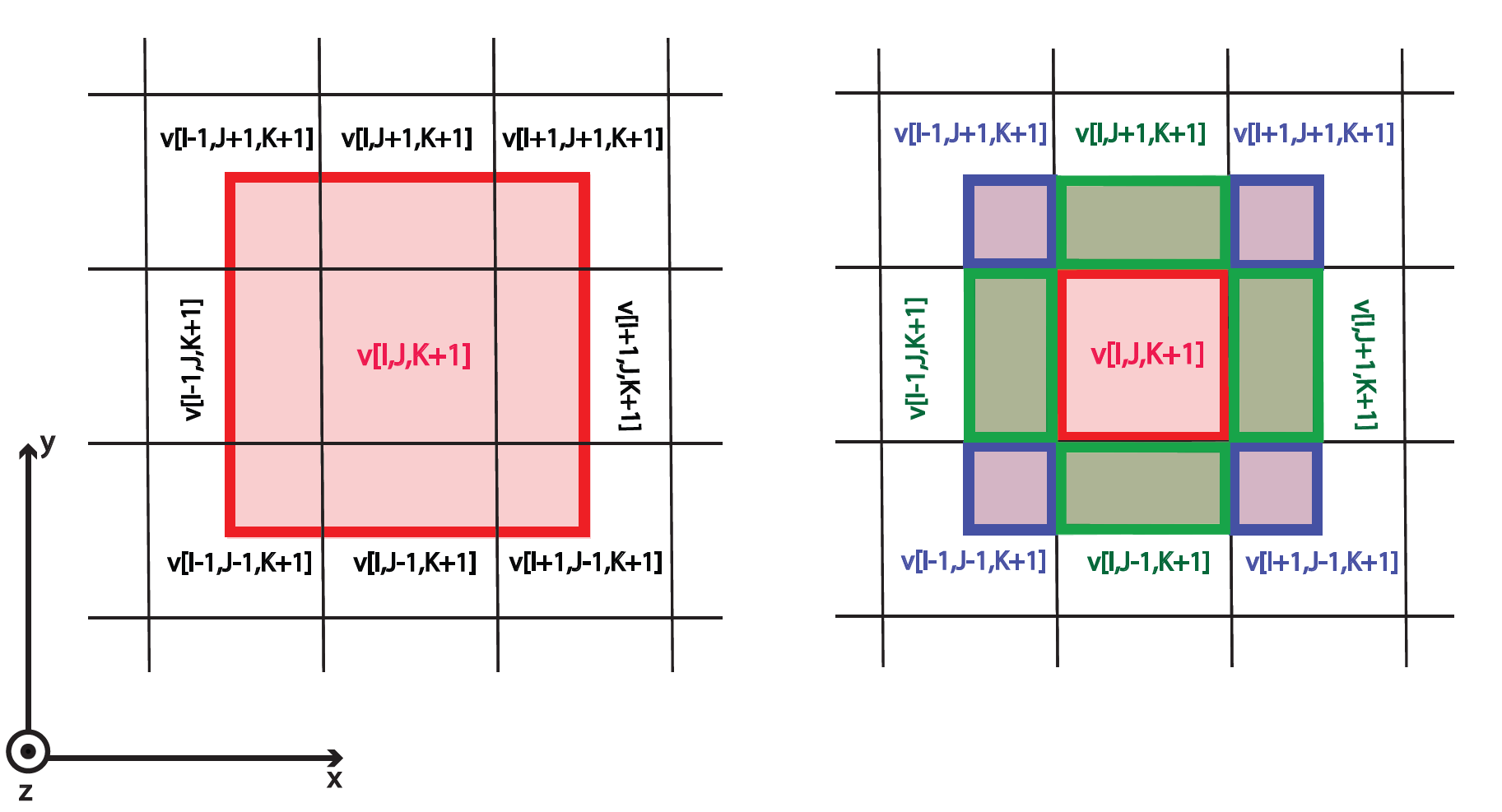}
    \caption{Graphic representation of the upper face of a box with side $2s$ enclosing the pixel $[I,J,K]$. The finite difference method approximate the flux across this surface as simply the contribution of $\mathbf{v}[I,J,K+1]$ over it (left). Meanwhile, the integral approximation technique considers the contributions of the side and of the diagonal pixels as well (right).}
    \label{Grid}
\end{figure*}


%

\section{Divergence reconstruction}\label{Meth}

\subsection{Finite differences}

We use the central finite difference method to compute derivatives in each pixel of the data-cube as:
\begin{eqnarray}
    \nabla \mathbf{v}=\partial_j v_i\approx\frac{v_i(x_j+s )-v_i(x_j-s)}{2s}\,, \label{finitedifference}
\end{eqnarray}
where $x_i$ is the central point of a pixel $(I,J,K)$ of the array. Also, $s$ is the physical size of a pixel, so we can use directly the cube data to ensure the right conversion of a pixel to the physical value, which fortunately is the same for each coordinate. For the case of the velocity field used here, with $257^3$ pixels in a datacube of length $400/h$ (where $h/100~km/secMpc$ is the dimensionless normalised Hubble parameter), we have:

\begin{eqnarray}
    s = \frac{400 Mpc}{257h}\,,
\end{eqnarray}

Neglecting the borders of the sample, leads to a $255^3$ array containing each pixel in a $3\times3$ matrix that corresponds to the gradient tensor of the peculiar velocity field. For a central finite difference approximation of a function $f$, one may write:
\begin{eqnarray}
\begin{split}
   f'(x) &= \frac{f(x+s)-f(x-s)}{2s}-\frac{s^2}{6}f'''(\xi)\\ &=f'_1(x)-\epsilon\,,
\end{split}
\end{eqnarray}
for some $\xi \in [x-s,x+s]$. In the above, $f'_1(x)$ is the central approximation for the derivative and $\epsilon \propto s^2f'''(\xi)$ is the truncation error. 



\subsection{Integral Approximations}

A direct approximation of the divergence could be computed recalling the definition of the operator:

\begin{equation}
        \nabla \cdot \mathbf{v}=\lim_{V\rightarrow 0} \frac{1}{V}\oiint_{\partial V}\mathbf{v}\cdot d\mathbf{A}\,.
\end{equation}
In so doing, we choose a box-like volume of size $2s$ around the central point of each pixel and compute the flux of the velocity field through the box. Then, by dividing the flux over the volume, we can extract an approximate value for the divergence. Note the difference with the central finite difference method in equation (\ref{finitedifference}), as this approach ignores the contribution over diagonal pixels (see Figure~\ref{Grid}) .

\subsection{Theoretical estimation}

Following equation (\ref{velrec}) the divergence of the velocity field and the density contrast are also related by the linear expression:
\begin{equation}
    \nabla \cdot \mathbf{v}(\mathbf{r})=\frac{\beta}{4\pi}\int^{R_{max}}d^3\mathbf{r'}\delta_g(\mathbf{r'})\nabla \cdot \frac{\mathbf{r'}-\mathbf{r}}{     \lvert \mathbf{r'}-\mathbf{r}\rvert^3 }=-\beta \delta_g(\mathbf{r})\,.  \label{exactdiv}
\end{equation}
Note that, as the $\mathbf{r}$ coordinate is measured in $km/s$, to express the divergence in $\frac{km/s}{Mpc/h}$ units, we need to multiply this quantity by a factor of $100h^2$. 

\section{Results}\label{Res}

We have computed the gradient matrix using the \texttt{Numpy} package from \texttt{Python} to manipulate matrix and arrays. The following three methods of estimating the volume scalar have been used:
\begin{enumerate}
    \item A decomposition of the full gradient tensor of the velocity field employing finite differences.
     \item A integration approximation for the divergence using a box volume around the pixels.
     \item A theoretical estimation by means of relation (\ref{exactdiv}).
\end{enumerate}

\begin{figure*}[ht]
    \centering
    \includegraphics[width=7.5cm]{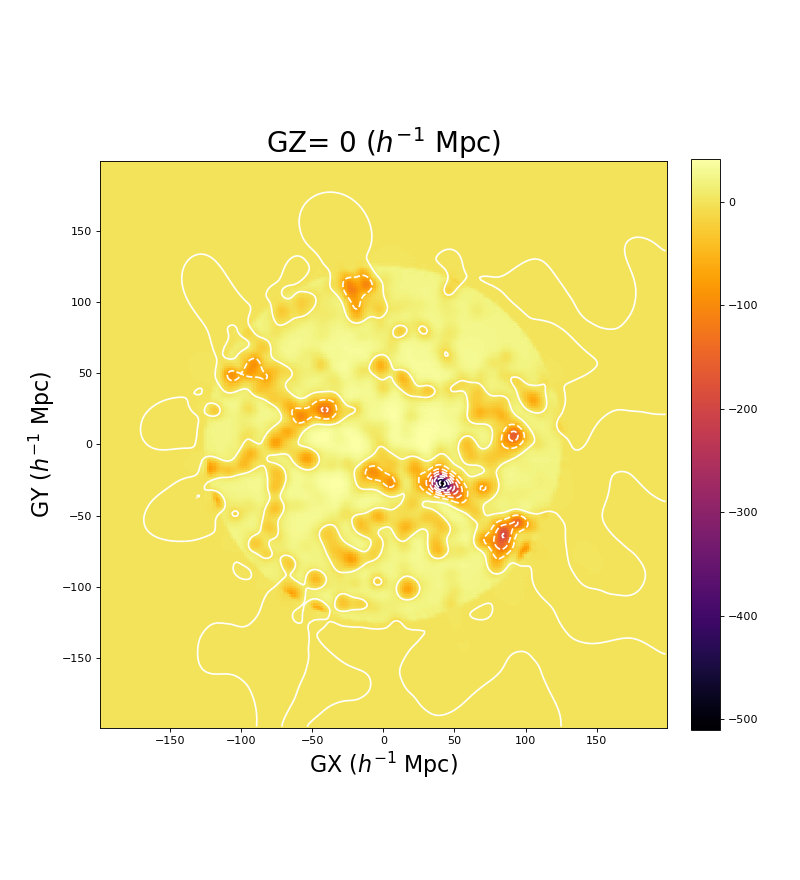}
    \includegraphics[width=7.5cm]{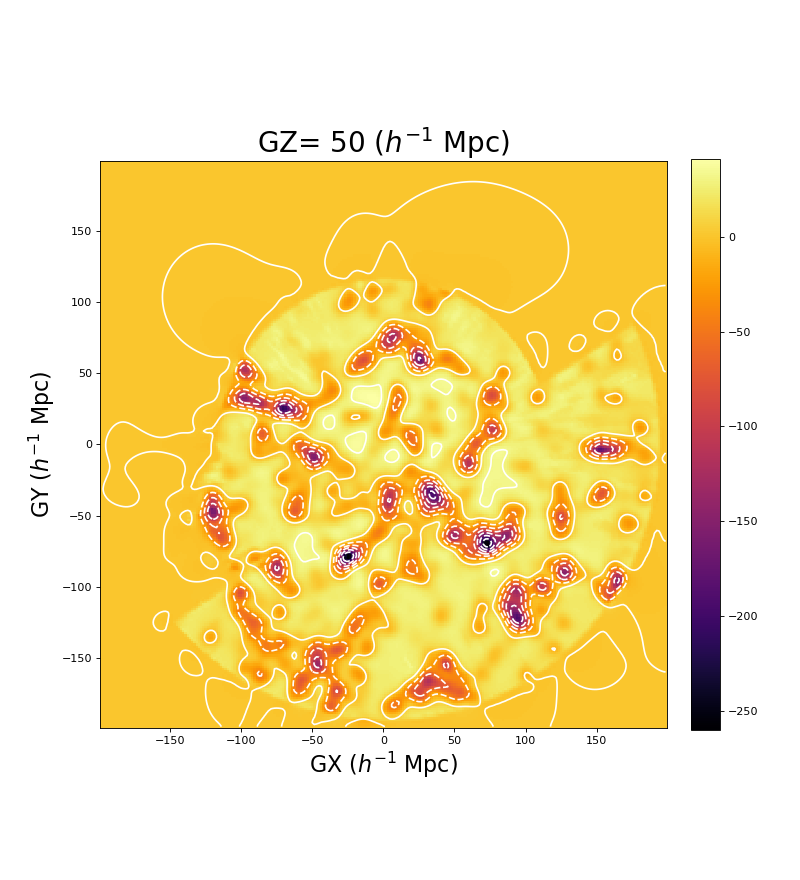}
    \includegraphics[width=7.5cm]{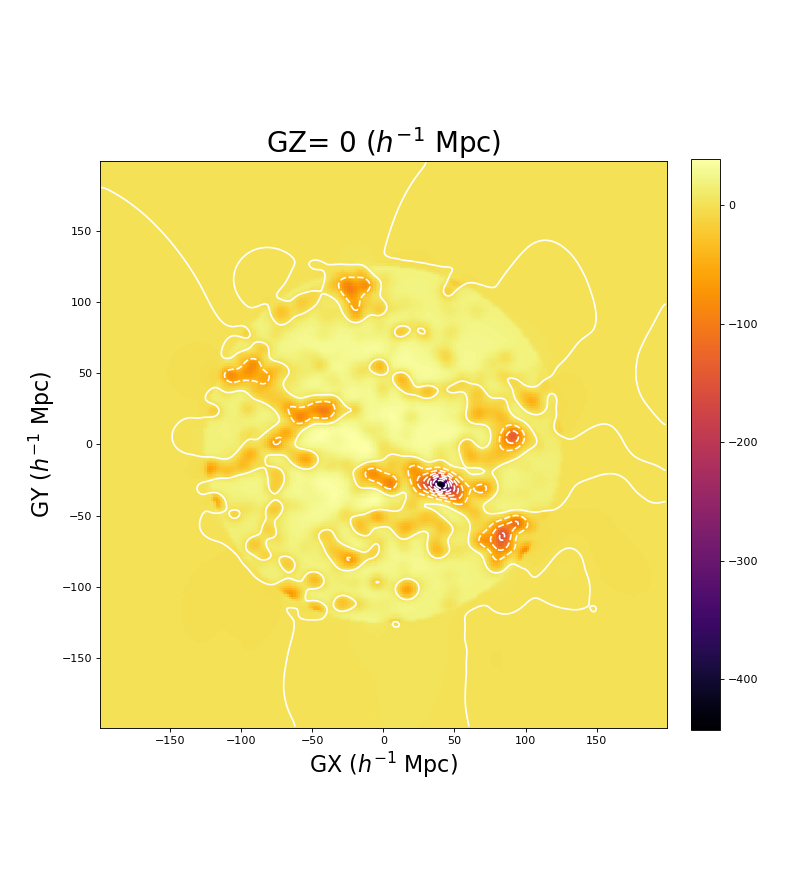}
    \includegraphics[width=7.5cm]{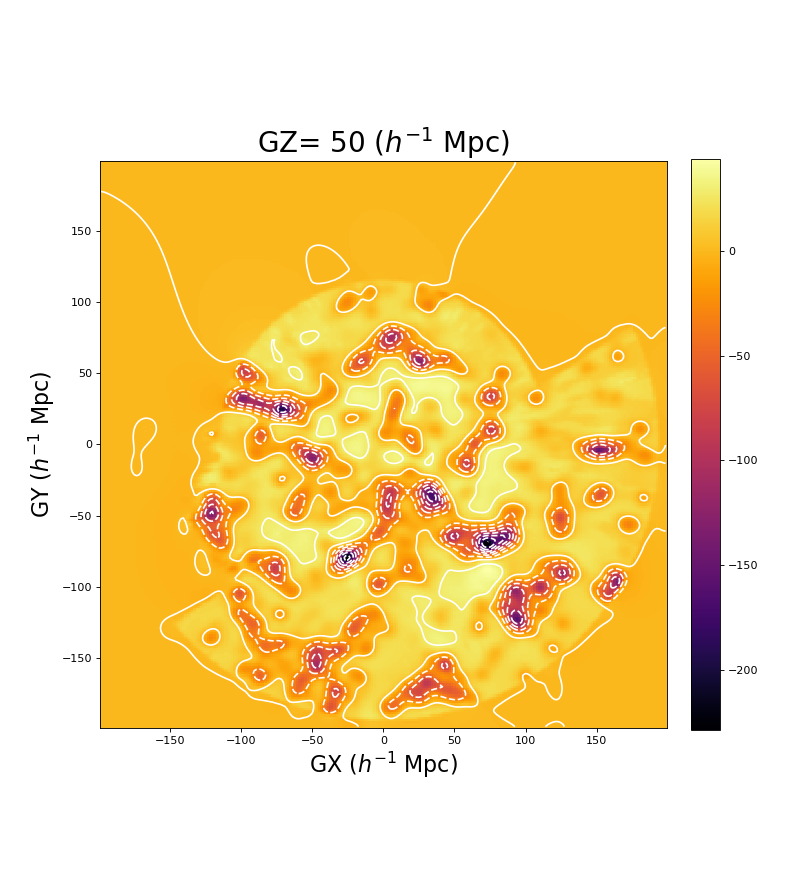}
    \includegraphics[width=7.5cm]{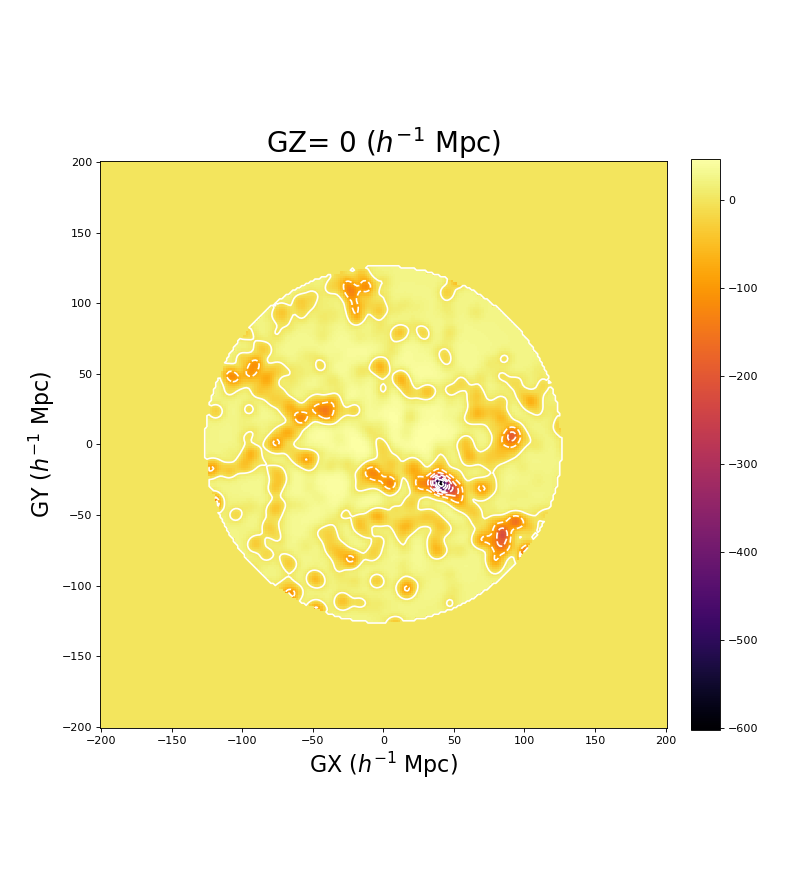}
    \includegraphics[width=7.5cm]{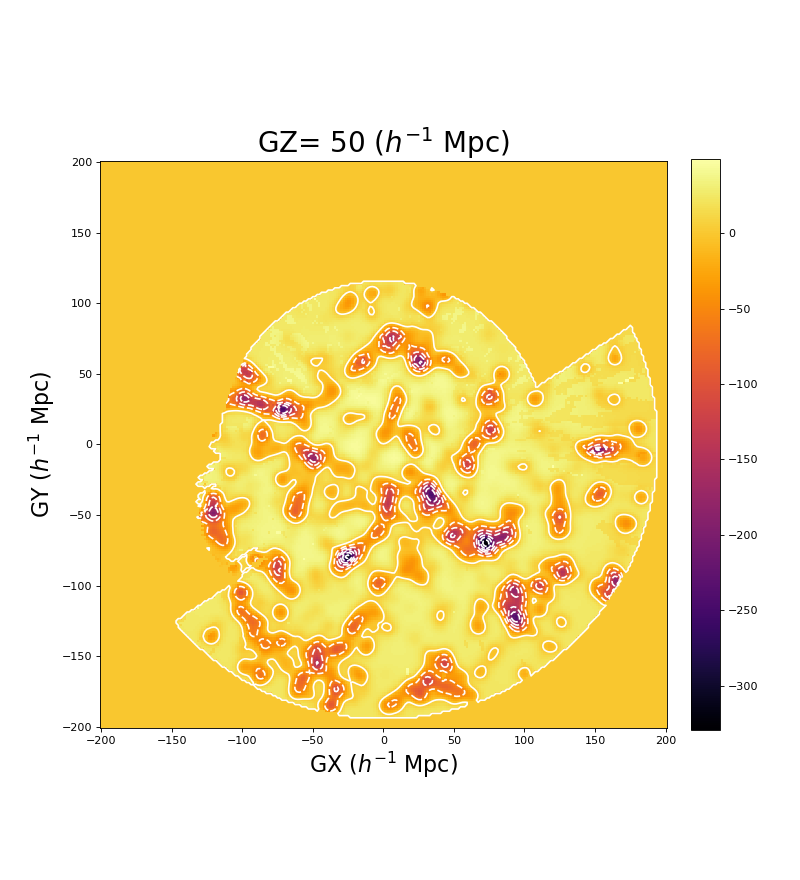}
    \caption{Divergence of the velocity field  in $\frac{km/s}{Mpc/h}$ using the finite difference, integral approximation and theoretical computation, projected over $GZ=0$ and $GZ=50 h^{-1}Mpc$ galactic planes.}
    \label{Divergence}
\end{figure*}


\begin{figure*}[ht]
    \centering
    \includegraphics[width=9cm]{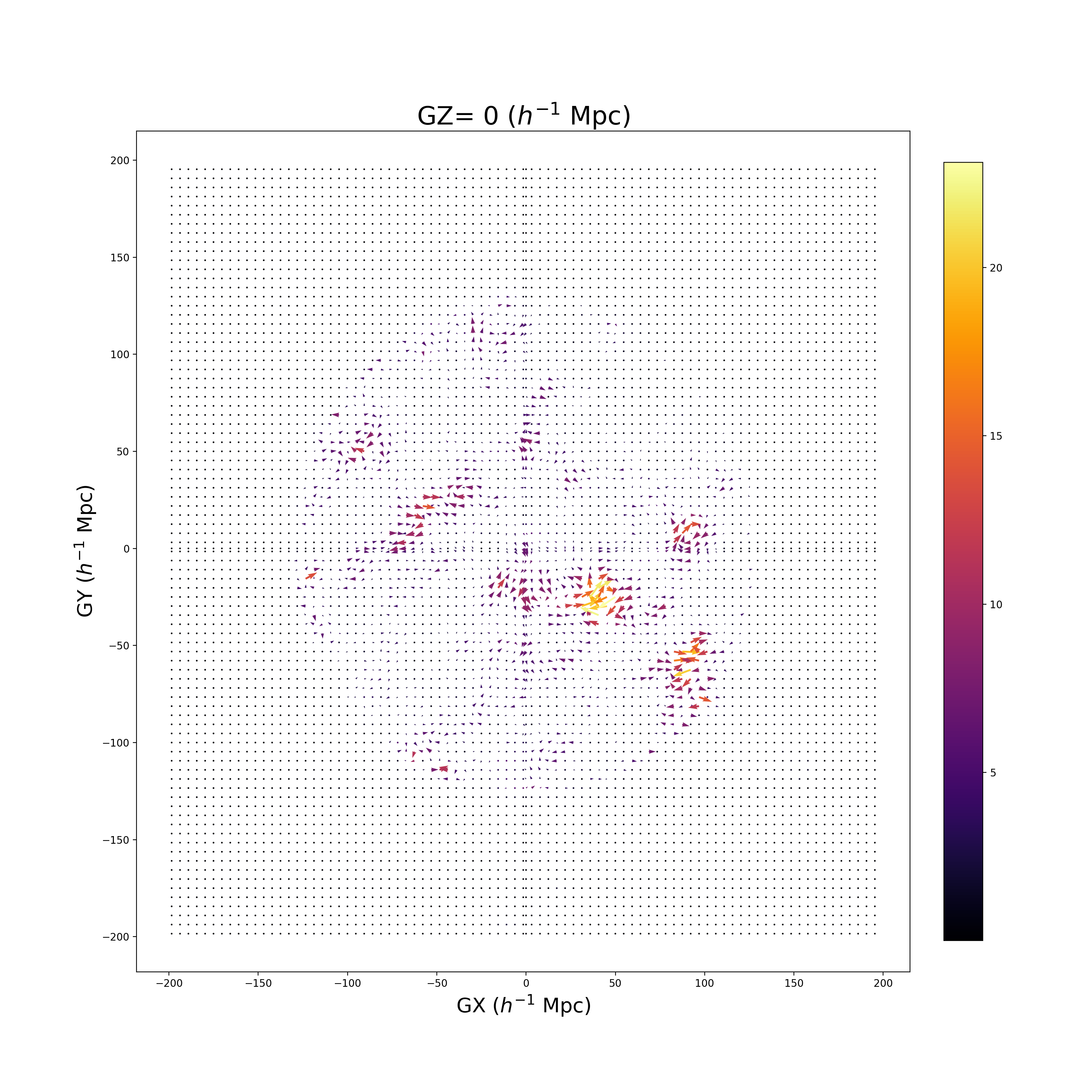}
    \includegraphics[width=9cm]{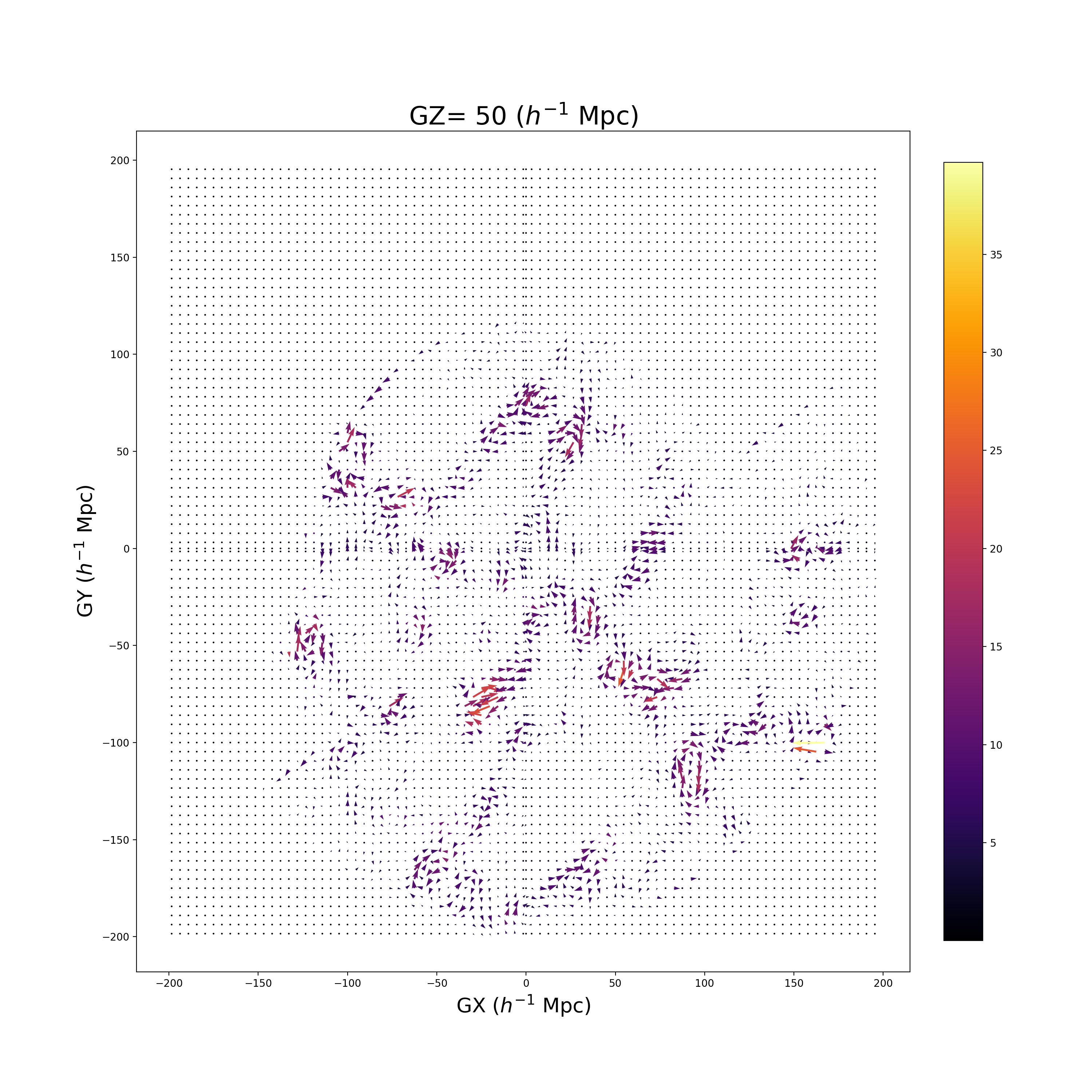}
    \caption{Residual curl vector field in $\frac{km/s}{Mpc/h}$ units from finite difference method, projected over $GZ=0$ and $GZ=50 h^{-1}Mpc$}
    \label{CurlVec}
\end{figure*}

\begin{figure*}[ht]
    \centering
    \includegraphics[width=8.5cm]{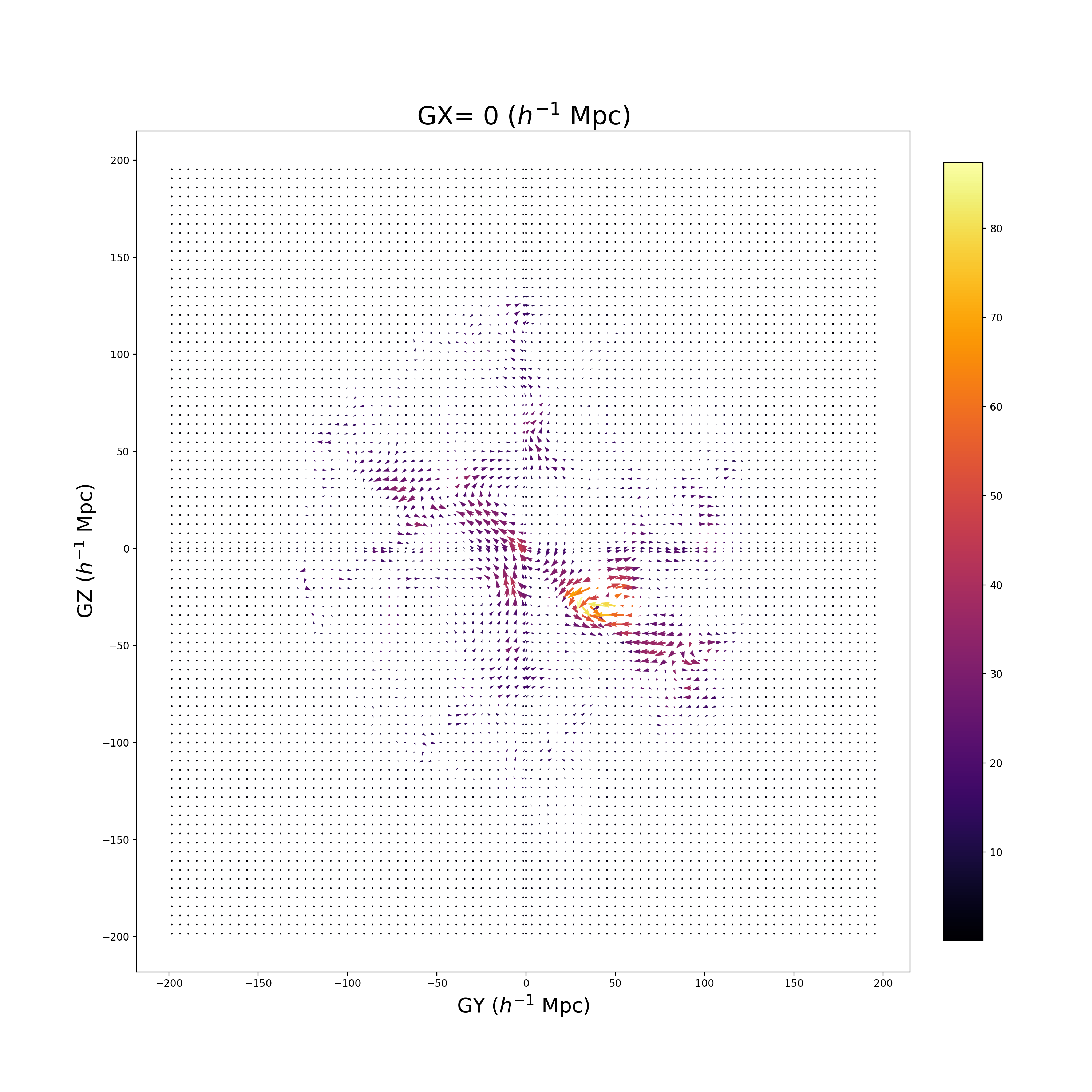}
    \includegraphics[width=8.5cm]{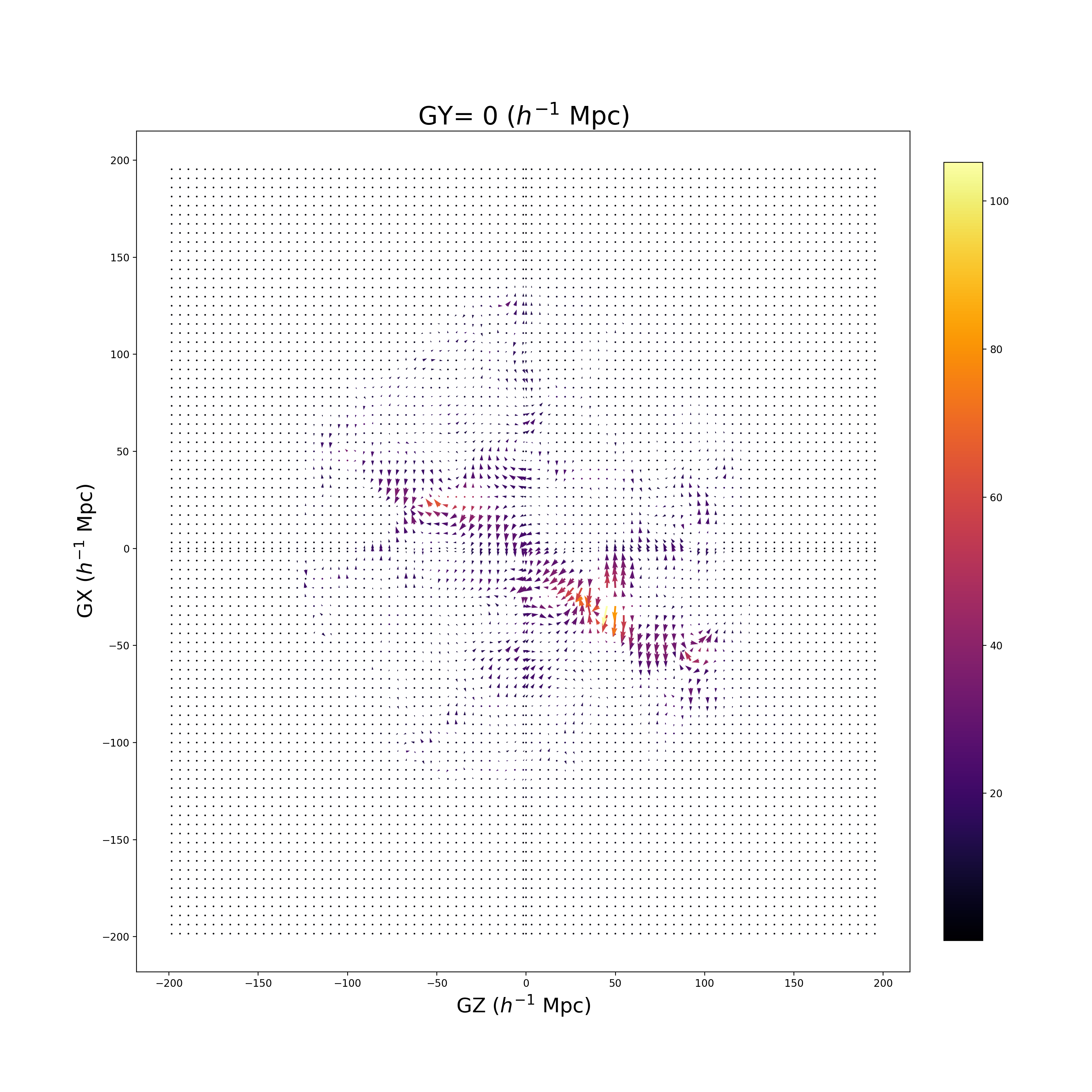}
    \includegraphics[width=8.5cm]{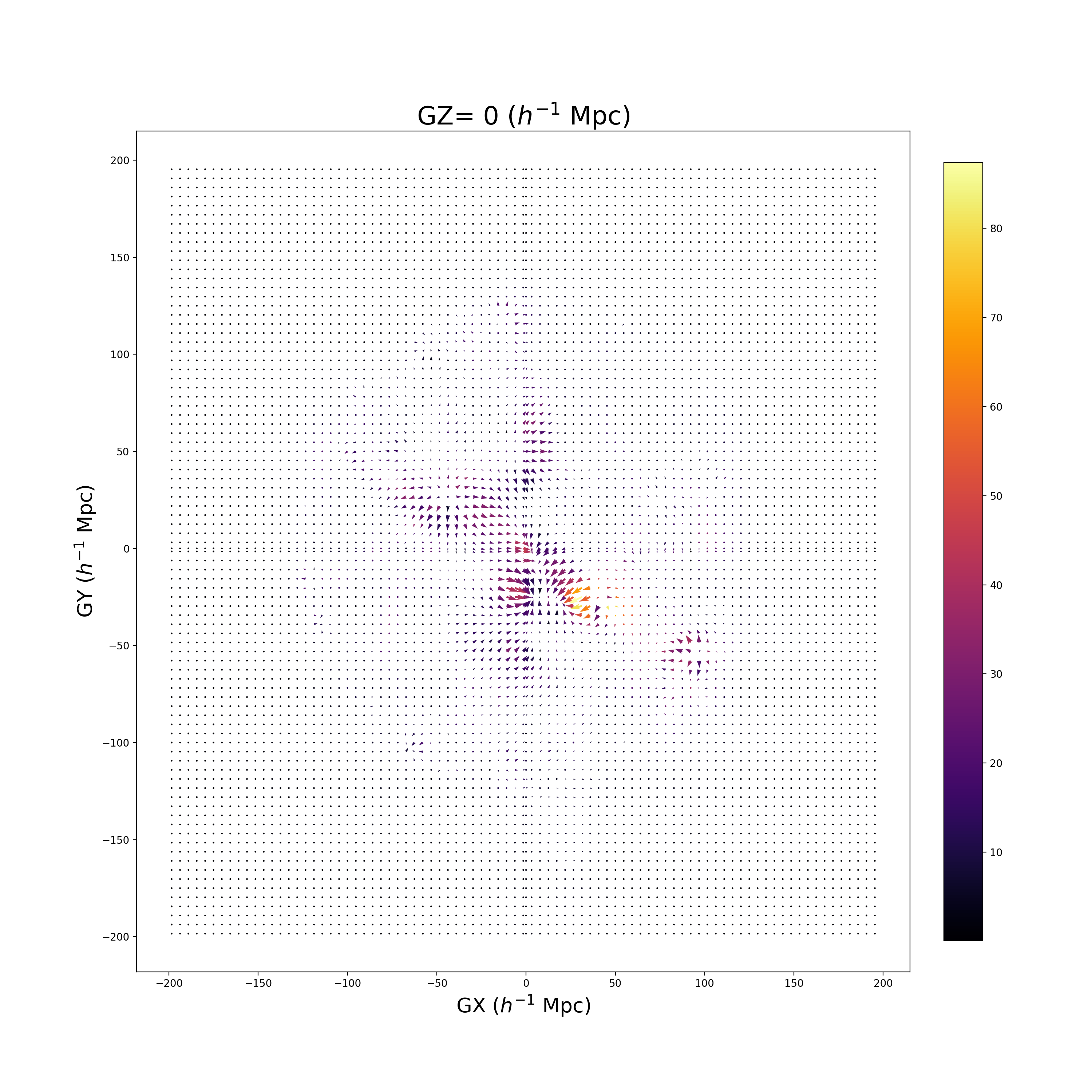}
    \caption{Projections of the shear tensor in $\frac{km/s}{Mpc/h}$ units estimated with finite difference method over cartesian planes that pass on the origin.}
    \label{Shear}
\end{figure*}

The results obtained via these different methods are plotted in Figure~\ref{Divergence}. To relate the latter with the tilted cosmology scenario, we need to estimate an average value for $\tilde{\theta}$ and thus put the predictions of the tilted model to the test. In our analysis, this corresponds to an average divergence of the entire fluid, which is then averaged over a spherical volume $V=4\pi\lambda^3/3$ as:
\begin{equation}
    \Tilde{\theta} = \frac{1}{V} \int_{V}(\nabla \cdot \mathbf{v}) dV \approx \frac{s^3}{V} \sum_{i}(\nabla \cdot \mathbf{v})_i\,,
\end{equation}
where the sum is over the pixels that reside inside a sphere of radius $\lambda$. 

Substituting $\tilde{\theta}$ into the right-hand side of equation~(\ref{tq2}), we can compute representative estimates of the local deceleration parameter ($\Tilde{q}$) measured by the bulk-flow observers on different scales ($\lambda$). The results, which assign negative values to $\tilde{q}$ on scales up to $200~Mpc$ through all three estimation methods, are summarized in Table~\ref{tabla1}. Note that we have set $h\simeq0.7$ in all cases. Also, although (\ref{tq2}) holds in essentially all tilted FRW models, here  we have assumed an Einstein-de Sitter background (with $q=0.5$) for mathematical simplicity. It seems that the theoretical estimation provides the higher values for $\Tilde{\theta}$, while the integral approximation gives the lowest. What is most important, however, is that all three methods are consistent both in the sign and in the magnitude of $\tilde{\theta}$. 

\subsection{Bias on the averaged density contrast}

According to equation (\ref{bias}), the difference between the mean density used to normalize the survey and the real mean density of the universe could introduce a significant bias in the results. If we express this relation in terms of divergence with a proper unit scaling, we obtain:

\begin{equation}
    \theta=(\bar{\delta}+1)\theta_s-100h^2f\bar{\delta}
\end{equation}

Where $\theta$ is the real divergence, $\theta_s$ is the divergence estimated from the reconstruction, and $f\sim \Omega_m^\gamma$ is the growth factor, which is approximately 1 for an EdS universe. Both divergences are in units of $\frac{km/s}{Mpc/h}$. A positive value of $\bar{\delta}$ indicates that the survey area coverage is inside an over-density region. In this case, the estimated divergence tends to be more negative, which supports the idea of a slightly contracting bulk flow. However, a negative $\bar{\delta}$ would increase the divergence to positive values and therefore should be carefully analyzed.

We can estimate the effects of this possible underdensity by assuming an extreme value for $\bar{\delta}$. For example, according to \cite{Haslbauer_2020}, for a scale $\sim 300 Mpc$ we can estimate the value of $\bar{\delta} \gtrsim -0.032$ from cosmic variance for a fiducial $\Lambda$CDM cosmology. The deviation $\Delta \theta$ for the estimated divergence from the real value is approximately:"

\begin{equation*}
    \Delta \theta \lesssim 1.6 \frac{Mpc}{h}
\end{equation*}

According to Table \ref{tabla1}, cosmic variance could strongly modify the values of the average divergence and even change its sign for scales $\lambda > 125 \frac{Mpc}{h}$. However, it's important to note that this is an extreme value of cosmic variance, and it's more likely that the deviation with respect to the mean density of the universe $\bar{\delta}$ is more positive than $-0.032$. Furthermore, a value of $\bar{\delta}$ that is much more negative than this cosmic variance could represent a problem for $\Lambda$CDM, as suggested in \cite{Haslbauer_2020}. Overall, we emphasize the necessity of broader surveys in order to better constrain the values of $\bar{\delta}$ and therefore $\theta$.

\subsection{Uncertainties}

We have identified both controlled and uncontrolled uncertainties in our estimations. In the former group we have the fit uncertainties for the reconstruction parameters $\beta$ and $\mathbf{V}_{ext}$. Of those two, we are mainly interested in $\beta$, given that $\mathbf{V}_{ext}$ does not enter the gradient calculation. Then, if we define the divergence of the relative velocity field $\mathbf{v}_{rec}$ as $\Tilde{\theta}_{rec}$, we have:
\begin{equation}
\Tilde{\theta}=\beta\Tilde{\theta}_{rec}\,,
\end{equation}
while the uncertainty in $\Tilde{\theta}$ due to the $\beta$ parameter can be written as:
\begin{equation}
\Delta\Tilde{\theta}_\beta=\Tilde{\theta}_{rec} \Delta\beta\,.
\end{equation}
This is the uncertainty recorded in Table~\ref{tabla1}. 

Turning to the uncontrolled uncertainties, we can group different possible systematic effects coming from the reconstruction process, as well as errors between approximations and real values. A detailed summary of the first type can be found in~\cite{Carrick_2015}. With regard to the approximation errors, we can estimate the precision of the estimation by comparing to the theoretical result. In this respect, the finite difference method seems more precise than the volume integration method, as it is closer to the theoretically predicted values. Moreover, according to relation (\ref{Reconstruction}), the velocity field should be irrotational as the field is proportional to a Newtonian gravity potential in the linear regime. However, when the anti-symmetric part of the gradient tensor is computed we got a non-zero value, leading to a residual low vorticity term that could be related with a deviation of the finite difference method with respect to theoretical estimation. (potential velocity) A symmetric trace-less part of the gradient can also be computed via finite difference method. Residual Curl and projections of the estimated Shear are plotted in Figures \ref{CurlVec} and \ref{Shear}.

\begin{table*}[ht]
    \centering
\begin{ruledtabular}
\begin{tabular}{@{}lcc@{}}
$\lambda(\frac{Mpc}{h})$ &$\Tilde{\theta}(\frac{km/s}{Mpc/h})$ & $\Tilde{q}$ \\
\hline
\midrule
 \textbf{Finite Difference}\\
 \hline
70  & $-3.36\substack{+0.07\\-0.07}\;(-2.65\substack{+0.08\\-0.12})$ & -6.36 (-4.90)\\
100  & $-2.77\substack{+0.06\\-0.06}\; (-2.19\substack{+0.10\\-0.07})$ & -2.27 (-1.70)\\
125  & $-1.48\substack{+0.03\\-0.03}\;(-1.17\substack{+0.06\\-0.04})$ & -0.45 (-0.25)\\
150  & $-0.65\substack{+0.01\\-0.01}\;(-0.51\substack{+0.02\\-0.02})$ & +0.21 (+0.27)\\
200  & $-0.24 \substack{+0.005\\-0.005}\;(-0.19\substack{+0.008\\-0.005})$ & +0.44 (+0.45)\\ 
\hline
\textbf{Integral Approximation}\\
\hline
70  & $-2.45\substack{+0.05\\-0.05}\;(-1.94\substack{+0.09\\-0.06})$ & -4.5 (-3.45)\\
100  & $-1.99\substack{+0.04\\-0.04}\;(-1.57\substack{+0.07\\-0.05})$ & -1.49 (-1.07)\\
125  & $-1.13\substack{+0.02\\-0.02}\;(-0.90\substack{+0.04\\-0.03})$ & -0.22 (-0.07)\\
150  & $-0.47\substack{+0.01\\-0.01}\;(-0.37\substack{+0.007\\-0.01})$ & +0.29 (+0.33)\\
200  & $-0.21 \substack{+0.004\\-0.004}\;(-0.17\substack{+0.008\\-0.005})$ & +0.45 (+0.46)\\
\hline
\textbf{Discrete Density Integration}\\
\hline
70  & $-3.94\substack{+0.08\\-0.08}\;(-3.11\substack{+0.15\\-0.10})$ & -7.54 (-5.86)\\
100  & $-3.17\substack{+0.07\\-0.07}\;(-2.50\substack{+0.12\\-0.08})$ & -2.67 (-2.00)\\
125  & $-1.66\substack{+0.03\\-0.03}\;(-1.31\substack{+0.06\\-0.04})$ & -0.56 (-0.34)\\
150  & $-0.76\substack{+0.02\\-0.02}\;(-0.59\substack{+0.03\\-0.02})$ & +0.16 (+0.23)\\
200  & $-0.29 \substack{+0.006\\-0.006}\;(-0.23\substack{+0.01\\-0.007})$ & +0.42 (+0.44)\\
\end{tabular}
\caption{Representative values for $\Tilde{q}$ on different scales ($\lambda$), using $\beta$ as it is in the datacube with the finite difference approximation, integral approximation and theoretical estimation. In parenthesis are the values of $\tilde{q}$ obtained after using $\beta$ from Pantheon+. Note that, for numerical simplicity and demonstration purposes, we have set $q=0.5$ in the CMB frame and $h\simeq0.7$.}
\label{tabla1}
\end{ruledtabular}
\end{table*}

\section{Discussion}
We have estimated the average volume scalar of the reconstructed peculiar velocity of the local universe via different methods. The volume scalar is related to the divergence of the velocity field. This is so because the velocity divergence measures the change in the local volume of the associated bulk flow and therefore its tendency to locally expand or contract. Then, a positive divergence implies that the fluid tends to expand locally, whereas a negative one indicates a contracting region. We have plotted the divergence scalar for different galactic planes in Figure \ref{Divergence}. There, one can see that the peculiar velocity divergence is highly negative in regions where the density contrast is high, while it is positive in regions where matter content is low. This is to be expected, of course, given the attractive nature of gravity. At this point, it also helps to recall the familiar divergence theorem:
\begin{equation}
        \oiiint_{V}(\nabla \cdot \mathbf{v}) dV=\oiint_{\partial V}\mathbf{v}\cdot d\mathbf{A}\,.
\end{equation}
Integrating the divergence over the region $V$ reveals whether the latter contracts or expands, as the right-hand side of the equation represents the fluid fraction that "enters" or "goes out" of the volume surface $\partial V$ over time. 

Surprisingly, the values of the local volume scalar ($\tilde{\theta}$) associated with the reconstructed peculiar velocity field, were found negative over a range of scales and by means of different estimation methods. This result has direct implications for the tilted cosmological scenario~\cite{Tsagas2011,Asvesta22},. The latter predicts that observers living in contracting bulk peculiar flows could measure a negative deceleration parameter locally, even when the universe is decelerating globally~\cite{Tsagas2015,Tsagas2021,Tsagas2022}. Also, as predicted, we found that the impact of the observer's peculiar motion becomes stronger on progressively smaller scales, namely closer to the observer, while it decays away from them (see Table~~\ref{tabla1}). The transition length ($\lambda_T$), that is the maximum scale where the local deceleration parameter appears to cross the $\tilde{q}=0$ mark and turn negative, also depends on the observer's position inside the bulk flow. Following (\ref{lambdaT}), for observes residing within $70/h$~Mpc from the centre of the bulk flow, we find $\lambda_T\gtrsim360$~Mpc, $\lambda_T\gtrsim310$~Mpc and $\lambda_T\gtrsim390$~Mpc, when adopting the Finite Difference method, the Integral Approximation method and the Discrete Density Integration method respectively. Overall, the closer the observer is to the bulk-flow centre, the more negative the local value of $\tilde{\vartheta}$ and the larger the associated transition length. 

As appealing these results may be, it is important to remain vigilant. It is possible, for example, that the values of the average divergence could change, as more refined surveys and models are developed. The value of the average divergence could be modified due to cosmic variance. which we estimate as $\Delta \theta\sim 1.6 \frac{Mpc}{h}$ for a fiducial $\Lambda$CDM cosmology. Recall that in the reconstruction used here this contribution was approximated by a constant velocity term. In addition, there have been recent claims that we live in a large void extending up to $\sim300~Mpc$. However, a negative expansion scalar is not compatible with the idea of a large void, where one expects to find an expanding bulk flow rather than a contracting one. In this respect, this velocity field reconstruction not seem to support the presence of a large underdensity. 

Finally, peculiar velocities seem unlikely to change the local value of the Hubble parameter appreciably and therefore to solve the $H_0$ tension. One can immediately realise this by looking at the linear relation (\ref{Thetas}a). Indeed, keeping in mind that $|\tilde{\theta}|/\Theta=|\tilde{\theta}|/3H\ll1$ on sufficiently large scales, the impact of the observer's relative motion on the Hubble parameter should be minimal.\footnote{Recall that, although $|\tilde{\theta}|/H\ll1$ always during the linear regime, this is not necessarily the case for the ratio $|\tilde{\theta}^{\prime}|/\dot{H}$.} Instead, there might be other explanations, such as systematics, the evolution of cosmological parameters with redshift, etc (e.g.~see~\cite{Krishnan_2020, Colgain22}).

\section{Conclusions}

We have computed the average divergence ($\Tilde{\theta}$) of the peculiar velocity field reconstructed from the $2M++$ survey, which was used to correct cosmological redshifts in the last SNIA compilation Pantheon+. In so doing, we employed three different approximation methods, coming from standard numerical analysis, the divergence theorem and from a linear theoretical derivation of the peculiar velocity formulae. In all cases, the resulting values of the velocity divergence were found negative over a range of scales, suggesting that we live inside a contracting bulk flow. According to the tilted cosmological scenario, the deceleration parameter measured locally by observers residing in contracting bulk flows can be negative, although the surrounding universe is globally decelerating. Our numerical results support this scenario, thus allowing for the recent accelerated expansion to be just an illusion produced by our peculiar motion relative to the CMB rest frame. Nevertheless, this possibility should be treated with care, as the computed values are still representative of the measurements a typical bulk-flow observer will make. Also, possible bias in the survey due to cosmic variance could be important to modify the value and even the sign of the average divergence over some large scales. Therefore, better surveys with refined precision and broader range are needed to improve the values computed here. In any case, however, our results support the need for a deeper study and for the proper understanding of the implications the observed large-scale peculiar motions may have for our interpretation of the cosmological parameters,

\section*{ACKNOWLEDGMENTS}
EP acknowledges support from the graduate scholarship ANID-Subdirecci\'on de Capital Humano/Doctorado Nacional/2021-21210824. 
We also wish to thank Christos Tsagas for his comments, which helped us understand further the tilted cosmological scenario.

\section*{DATA AVAILABILITY}
The data underlying this article, including the programs and the results of gradient estimations, will be shared on reasonable request to the corresponding author.

\bibliography{mybib}

\end{document}